\documentclass[reprint,preprintnumbers,amsmath,amssymb,aps,prl]{revtex4-1}
\usepackage{xargs}                      
\usepackage[pdftex,dvipsnames]{xcolor}  
\usepackage[colorinlistoftodos,prependcaption,textsize=tiny]{todonotes}
\newcommandx{\unsure}[2][1=]{\todo[linecolor=red,backgroundcolor=red!25,bordercolor=red,#1]{#2}}
\newcommandx{\change}[2][1=]{\todo[linecolor=blue,backgroundcolor=blue!25,bordercolor=blue,#1]{#2}}
\newcommandx{\info}[2][1=]{\todo[linecolor=OliveGreen,backgroundcolor=OliveGreen!25,bordercolor=OliveGreen,#1]{#2}}
\newcommandx{\improvement}[2][1=]{\todo[linecolor=Plum,backgroundcolor=Plum!25,bordercolor=Plum,#1]{#2}}
\newcommandx{\thiswillnotshow}[2][1=]{\todo[disable,#1]{#2}}
\usepackage{graphicx}
\usepackage{dcolumn}
\usepackage{bm}
\usepackage{algorithm}
\usepackage{algpseudocode}

\usepackage{multirow}

\newcommand{\GM}[1]{$\mathcal{G}$({#1})}
\newcommand{\ucbMatsci}{Department of Material Science and Engineering, University of California, Berkeley, CA 94720, U.S.A.} 

\newcommand{\LBLMSD}{Materials Sciences Division, Lawrence Berkeley  National Laboratory, Berkeley, 94720, United States}

\begin{document}

\title{Topological Graph-based Analysis of Solid-State Ion Migration}

\author{Jimmy-Xuan Shen}
\affiliation{\ucbMatsci}
\author{Haoming Howard Li}
\affiliation{\ucbMatsci}
\author{Ann Rutt}
\affiliation{\ucbMatsci}
\author{Matthew K. Horton}
\affiliation{\LBLMSD}
\affiliation{\ucbMatsci}
\author{Kristin A. Persson}
\affiliation{\LBLMSD}
\affiliation{\ucbMatsci}

\date{\today}

\begin{abstract}
    To accelerate the development of novel ion conducting materials, we present a general graph-theoretic analysis framework for ion migration in any crystalline structure. The nodes of the graph represent metastable sites of the migrating ion and the edges represent discrete migration events between adjacent sites. Starting from a collection of possible metastable migration sites, the framework assigns a weight to the edges by calculating the individual migration energy barriers between those sites. Connected pathways in the periodic simulation cell corresponding to macroscopic ion migration are identified by searching for the lowest-cost cycle in the periodic migration graph. To exemplify the utility of the framework, we present the automatic analyses of Li migration in different polymorphs of VO(PO$_4$), with the resulting identification of two distinct crystal structures with simple migration pathways demonstrating overall $< 300$~meV migration barriers.
\end{abstract}


\maketitle


\section{Introduction}\label{sec:introduction}

The migration of charged ions ({\it eg.} Li, Mg, Na, O$^{2-}${\it etc}.) through solid-state materials is the primary physical mechanism behind the operation of Li-ion batteries, solid-oxide fuel cells, and solid-state electrolytes.
Rapid identification and discovery of new materials with favorable migration characteristics is key to developing all-solid-state batteries where the current state-of-the-art organic electrolytes are replaced with a solid-state alternative, leading to improved power density and safety.
Traditionally, the discovery of novel electrode materials has focused on compounds that contain the migrating ion in their as-synthesized state.
However, this is not a strict requirement, and many materials synthesized without the migrating species are capable ion conductors. 
In fact, it has been shown that for multivalent applications, materials that are synthesized without the working ion tend to exhibit a flatter migration energy landscape and hence better performance ~\cite{Kim2020Oct,Sun2016Jul,Rong2015Sep,Levi2010Jun}.

The established method for identifying the optimal path between two sites in a crystal is the nudged-elastic band (NEB)  method~\cite{Mills1994Feb,Jonsson1998Jun}.
However, NEB calculations are computationally costly and are only able to analyze short-distance migration events provided that an initial, reasonably accurate, guess for the connecting path is available.
To understand the migration characteristics of a material, the motion of the ion through the entire crystal must be considered. 
Recent high-throughput studies have attempted to address this either by simplifying the problem to analyzing the migration of a working ion in a fictitious field~\cite{Kahle2020Jan} or by focusing on individual migration events but not how they connect over larger distances~\cite{Bolle2020Jun}.
Additionally, previous work exclusively treat materials where valid sites for the working ion are known beforehand.
To explore the broader class of materials, where there is no {\it a priori} knowledge of the sites and migration properties of the possible intercalants, it is of considerable interest to develop algorithms and frameworks to analyze possible ion migration behavior in any crystalline solid. 

In this endeavor, we employ a recently developed methodology where the charge density analysis was shown to be a reliable descriptor for generating initial guesses of working ion sites~\cite{Shen2020Oct} which allows us to systematically identify metastable intercalation sites in any crystalline structure. Here, we build upon this framework and present a graph theory extension to automatically identify ion migration pathways in any periodic solid.
The migration is treated as a periodic graph where symmetrically equivalent copies of the metastable sites constitute the nodes and the individual migration events between these sites are the edges.
Additionally, we assign a cost to the graph edges based on the migration energy barriers and showcase how optimal intercalation pathways can be discovered with a Dijkstra's-inspired algorithm defined on the periodic graph.
The original code provided here is distributed as an extension to the \texttt{pymatgen} material analysis library.
We demonstrate our framework on two well-known structures of MnO$_2$ and CoO$_2$ to show how the migration graphs can be constructed and utilized.
Finally, the methodology is applied to the different configurations of VO(PO)$_4$ in the Materials Project~\cite{Jain2013Jul} to assess the migration characteristics of each polymorph, and we exemplify the capability to identify promising new ionic conductors within this set of materials.


\section{Results \& Discussions}\label{sec:results}

\subsection{Site identification}

Our graph-based migration analysis is best suited for the two limits of working ion occupation, either single-ion migration in the dilute limit or vacancy migration in the fully intercalated limit.
While it is possible to analyze intermediate concentrations, the large configurational space associated with the working ion ordering arrangements make a thorough investigation computationally demanding and not suitable for high-throughput evaluation of viable intercalation pathways.
For vacancy migration, {\it a priori} knowledge of the working ion sites makes the construction of migration graphs trivial.
In materials where we lack knowledge of working ion sites, we utilize a recently developed, generally robust computational workflow for identifying the metastable sites of the working ion in any structure.~\cite{Shen2020Oct} 
The methodology selects sites at the local minima of charge density and, for each candidate site, a working ion is inserted and the structure is allowed to relax using density-functional theory calculations. An inserted structure is considered ``topotactic'' if the positions of its framework atoms closely resemble the relaxed atomic positions of the host material.
The metastable sites are obtained by mapping the working ion in the topotactically inserted structures onto the empty host structure and identifying all symmetry-equivalent positions in the host structure. Based on the location and connectivity of the metastable sites, we build our graph-based migration analyses. 

\begin{figure*}[htb]
    \centering
    \includegraphics[width=0.95\textwidth]{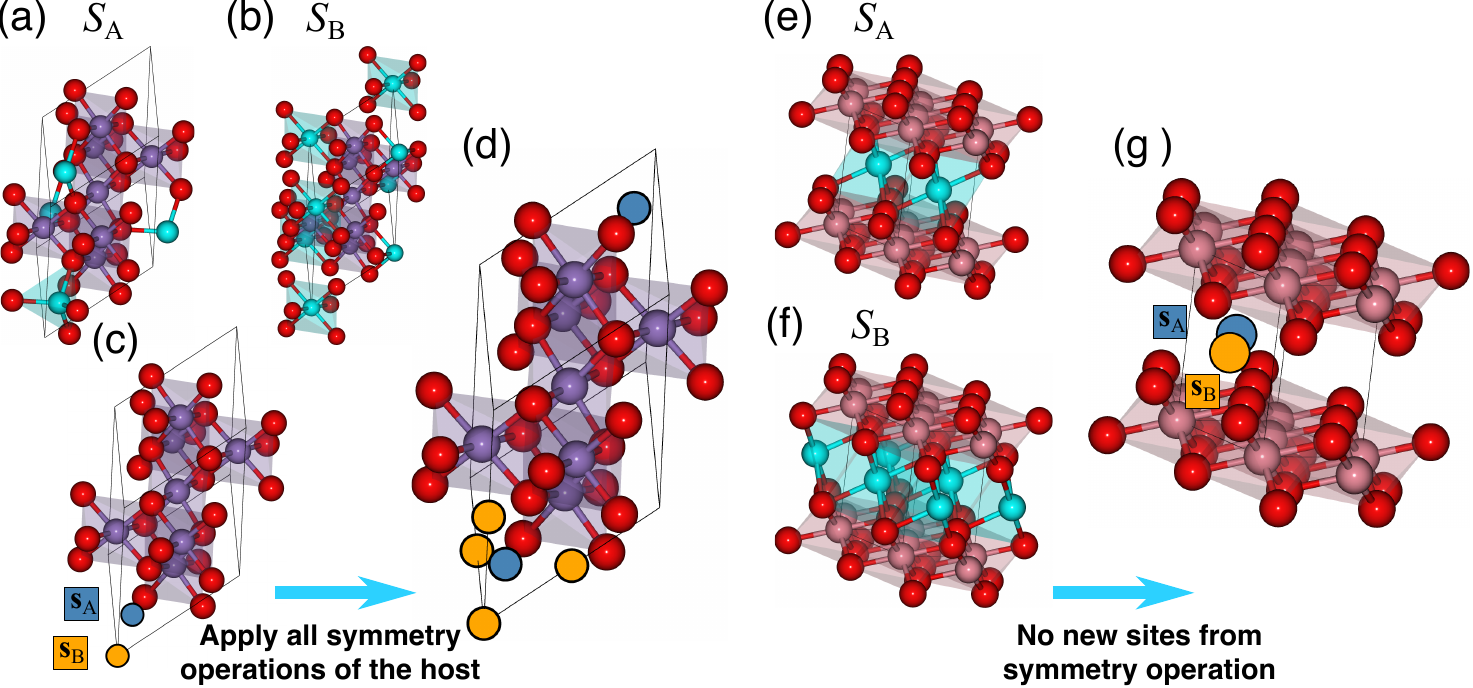}
    \caption{
        \label{fig:gen_meta_stable}
        Illustration of identified metastable sites in (a-d) MnO$_2$ and (e-g) CoO$_2$. (a-b) Crystal structure indicating the different relaxed atomic structures of MnO$_2$ after single Li insertion. (e-f) Crystal structure indicating the different relaxed atomic structures of CoO$_2$ after single Li insertion.
        (c) Crystal structure indicating the metastable Li position after mapping onto $S_{\rm base}$.
        (d,g) Crstyal structure with the full set of possible Li sites after symmetry operations. Note that the applied symmetry operations did not result in additional sites for CoO$_2$.
    }
\end{figure*}

To exemplify our approach, we use two materials: MnO2$_2$ in the $\lambda$ phase~\cite{Juran2018Apr} with cubic spinel structure and layered CoO$_2$ with ABBA stacking~\cite{Laubach2009Apr}.
After performing indepedent single Li insertions into the sites suggested by the charge density analysis and relaxing the new structures~\cite{Shen2020Oct}, two distinct singly-inserted structures for each material were topotactically matched to the host material as shown in Fig.~\ref{fig:gen_meta_stable}.
We denote the base structure $S_{\rm base}$ and the set of relaxed inserted topotactic structures $\{S_{\alpha}\}$ where $\alpha \in \{A, B\}$ for both examples.
Since the host sublattice (which does not contain the working ion) of each $S_\alpha$ can be mapped onto $S_{\rm base}$, the relaxed positions of the cations in each structure can also be mapped to position $\mathbf{s}_\alpha$ in $S_{\rm base}$.
This mapping allows us the identify two symmetry-distinct metastable sites $\mathbf{s}_A$~(blue) and $\mathbf{s}_B$~(orange) for MnO$_2$ and CoO$_2$, respectively. Utilizing the \texttt{spglib} package~\cite{Togo2018Aug} and its interface with \texttt{pymatgen}~\cite{Ong2015Feb}, we analyze the crystal symmetry of the structure with the inserted ion, $S_{\rm base}$, and apply the valid point group operations to each $\mathbf{s}_\alpha$ to generate all of the possible cation positions, designated by an integer index value $i$ at position $\mathbf{r}_i$, in the unit cell.

In MnO$_2$, the $\mathbf{s}_A$ metastable site is represented by the fractional coordinates $\left(\frac{1}{8},\frac{1}{8},\frac{1}{8}\right)$ and all space-group operations of the host material will either map the site to itself or $\left(\frac{7}{8},\frac{7}{8},\frac{7}{8}\right)$.
The $\mathbf{s}_B$ site is represented by the fractional coordinates $(0,0,0)$, which has three additional symmetry-equivalent sites as shown in Tab.~\ref{tab:li_positions}.
This results in a total of six metastable sites per unit cell as shown in Fig.~\ref{fig:gen_meta_stable}~(d).
We perform the same analysis for CoO$_2$, which results in $\mathbf{s}_A$ and $\mathbf{s}_B$ at the face centers of the primitive cell.
The space-group operations of CoO$_2$ map the sites onto periodic images of the original, as such, no new symmetrically equivalent sites are created from symmetry operations, the resulting two metastable sties are shown in Fig~\ref{fig:gen_meta_stable}~(g).
{}


\subsection{Graph Analysis}

Using a distance cutoff of $l_{\rm max}$, we connect two nearby metastable sites $\mathbf{r}_i$ and $\mathbf{r}_j$ to represent a discrete migration event in the material which we will call a ``{\it hop}''.
The network formed by these hops is infinite and the following convention ensures that we only consider hops that are inequivalent by lattice transitions.
Each hop between sites $i$ and $j$ in the periodic unit cell is labeled $h_{ij}^{\mathbf{K}}$ where the additional index $\mathbf{K}$ is an integer-valued vector representing the relative periodic image displacement between the endpoints [ie. $\mathbf{K} = (0,0,1)$ means that the hop crosses a period cell boundary once in the positive c-direction].
In general, we consider the migration graph to be undirected. 
As such, the hops $h_{ij}^{\mathbf{K}}$ and $h_{ji}^{-\mathbf{K}}$ will represent the same migration event, but only one representation is needed.
As a convention to prevent double-counting, we require the site indices to satisfy $i \leq j$.
Additionally, since there is ambiguity when $j = i$ and $\bm{K}\neq 0$, we only retain the hop where the first non-zero component of $\mathbf{K}$ is positive.

\begin{figure}[htb]
    \centering
    \includegraphics[width=0.5\textwidth]{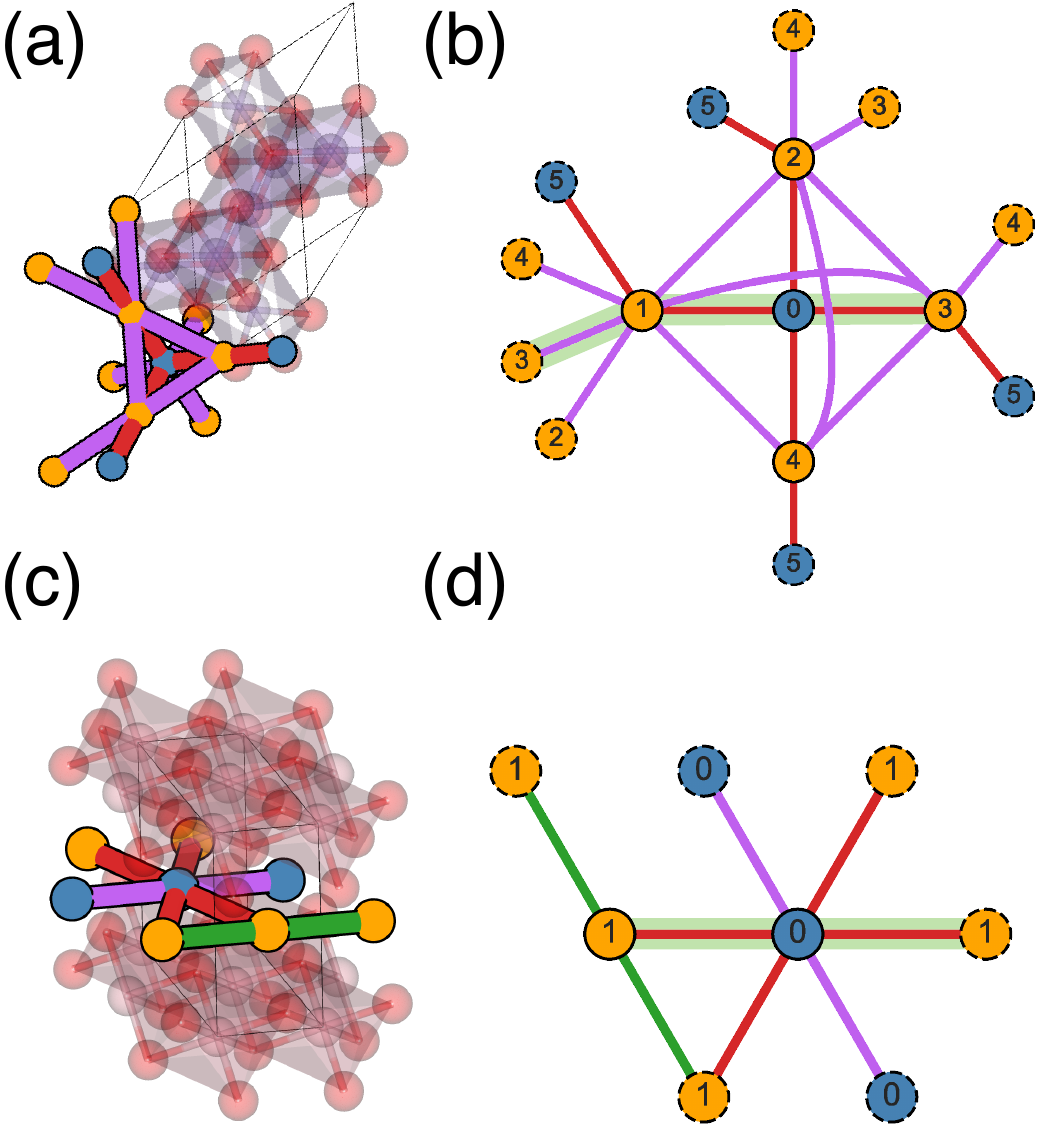}
    \caption{
        \label{fig:graph_rep}
        Graph representation of the migration hops in MnO$2$ (a,b) and CoO$_2$ (c,d).
        (a,c) Crystal structure with hops shown that fall below 3~\AA{} and are colored by symmetric equivalence.
        (c,d) The migration graphs for the two materials MnO$2$ and CoO$_2$ are shown, where
        dashed nodes represent $j$-index nodes that are outside the (0,0,0) unit cell.
        Although there is only one periodic copy of each node, the dashed nodes are used to differentiate the multiple edges connecting the same two nodes.
        Two examples of intercalating pathways are high-lighted in light-green in (b) and (d).
    }
\end{figure}

Using a threshold value of $l_{\rm max} = 3$~\AA, the migration graph for $Li^{+}$ in MnO$_2$ (denoted as \GM{MnO$_2$}) is constructed and shown in Fig.~\ref{fig:graph_rep}~(a-b).
There are 18 hops in \GM{MnO$_2$} that are not equivalent under discrete lattice translations.
Using the space group symmetry between the hops, we can reduce them to 2 symmetry-distinct groups indicated by the edge color in the graph.
The $Li^{+}$ migration graph in CoO$_2$, with 8 hops in 3 symmetry-distinct groups, is shown in Fig.~\ref{fig:graph_rep}~(c,d).
For a complete enumeration of the migration hops in these two materials and their symmetry equivalence, see Table S.II and S.III of the supplemental materials~\cite{Supp} (SI).
In principle, once we have identified the symmetrically equivalent groups, we can obtain the migration barrier using nudged elastic band (NEB) calculations~\cite{Jonsson1998Jun} to chart the migration energy landscape of the material.
%
{}

A candidate ion-conducting material must enable a continuous migration pathway for the working ion across the unit cell, connecting to the next one. In a periodic system, continuous pathways are infinite, which we term ``{\it intercalating pathways}''.
Since our migration graph contains only one copy of each node, periodicity manifests via the image displacement vector $\mathbf{K}$. 
The intercalating pathways are essentially cycles in the graph where the total image displacement is non-zero.
The series of hops in such a cycle will connect a metastable site to a different periodic image of itself, which constitutes a repeating unit of an infinite periodic migration pathway.
Basic examples of intercalating pathways are highlighted in light green in Fig.~\ref{fig:graph_rep} (b) and (d), which connect a node to a periodic image of itself.
To identify these pathways, we used a modified Dijkstra's type algorithm on the periodic graph. 
The key difference between the modified algorithm and the original Dijkstra's algorithm is that the periodic image vector is tracked during graph traversal. This means that the optimal cost to reach any node during the graph traversal is defined for the combination of node index $i$ and periodic image vector $\mathbf{K}$. A detailed description of the path-finding algorithm on the periodic graph is presented in the SI; Algo.~S\ref{algo:periodic}.
The cost function employed in the path-finding algorithm can be any positive definite function assigned to the edges of the graph. A good choice in most cases is the migration energy barrier for the ion-migration event represented by that particular edge.  However, the difference between the binding energies of the endpoints, which can be computed without expensive NEB calculations, may also be used as a lower bound of the activation barrier for screening purposes.

\subsection{Application to Polymorphs of VO(PO$_4$)}
We demonstrate the utility of the obtained migration graphs for Li migration in VO(PO$_4$).
Of all 18 VO(PO$_4$) phases currently available in the Materials Project, five are distinct known, synthesized phases with the following IDs(spacegroup symbols):
mp-25265(Pnma), 
mp-556459(Cc), 
mp-559299(P4/n), 
mp-763482(P4/n),
mp-1104567(C2/m).
While all five phases listed above have been experimentally synthesized, only some of them have readily available electrochemical analysis data. In particular, mp-25265 ($\beta$-VOPO$_4$) demonstrates a capacity of 118.6 mAh/g against Li insertion at average 4V~\cite{Ren2009Apr}, and mp-556459 ($\varepsilon$-VOPO$_4$) has shown a capacity of 305 mAh/g against Li insertion over two voltage plateaus at about 4.0 and 2.5 V~\cite{Siu2018}.
For each of the five structures, we performed a set of ion insertions to generate metastable sites and constructed the migration graphs, yielding connectivity of hops to form intercalation pathways. With this connectivity, the only missing piece to a complete description of the intercalation behavior is understanding of the ion migration energy evolution during the hops.

The energy profile for each hop can be estimated by implementing the ApproxNEB method~\cite{Rong2016Aug} in \texttt{atomate}.
The ApproxNEB method performs independent constrained optimizations for each image structures which allows us to trade accuracy for speed since the independent relaxations are trivially parallelised.
Each phase of VO(PO$_4$) has 3 to 10 such hops and thus 3 to 10 ApproxNEB calculations. 
Due to the high computational cost involved, one might find it helpful, in general, to rank migration pathways before ApproxNEB is employed. 
To demonstrate testing of one possible choice of cost function for this purpose, we performed charge-density analyses on these phases and compared them to our ApproxNEB results.

\begin{figure}[htb]
    \centering
    \includegraphics[width=0.5\textwidth]{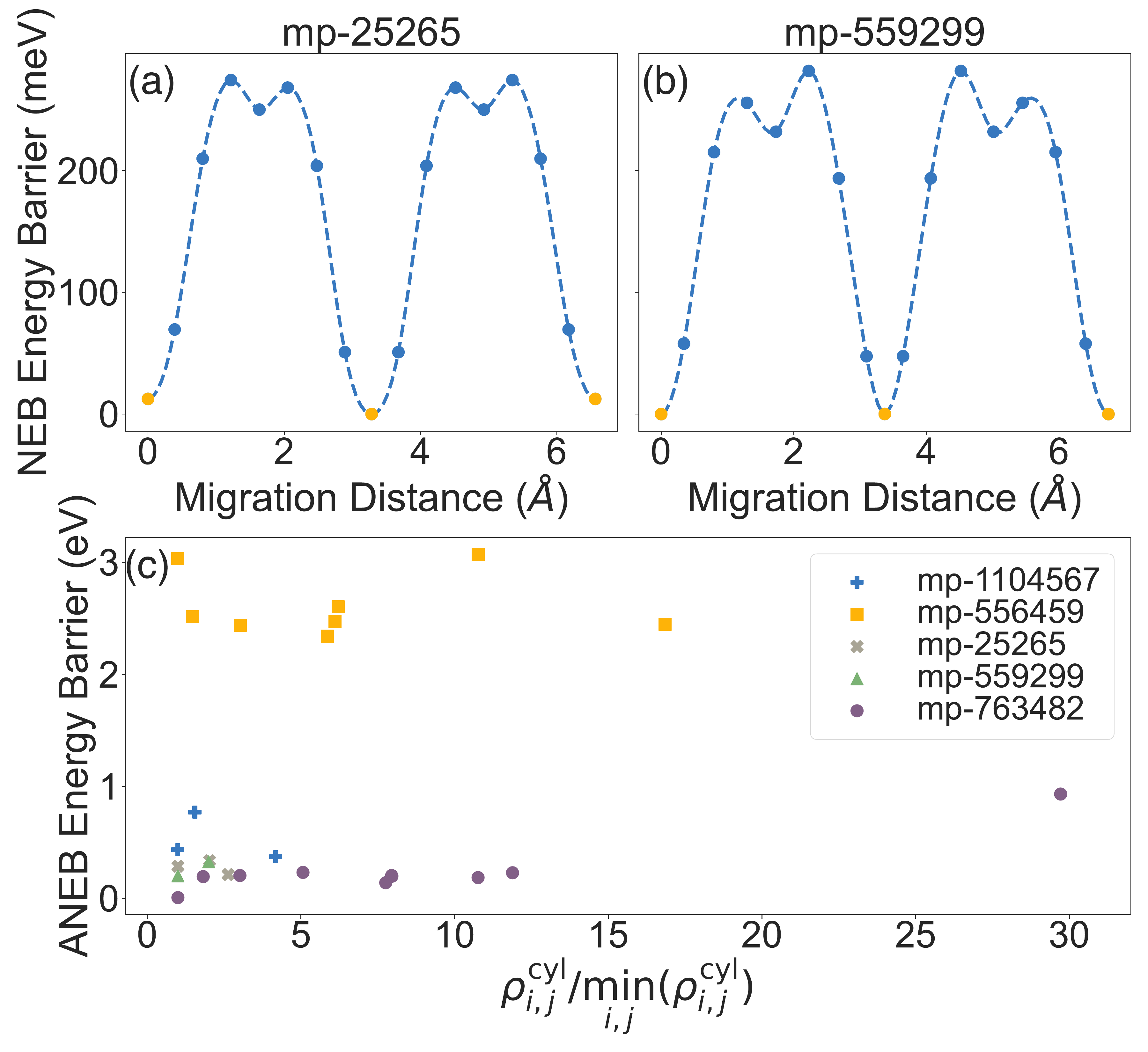}
    \caption{
        \label{fig:aneb} 
        The NEB-calculated energy landscapes along the lowest-barrier paths for mp-25265 (a) and mp-559299 (b) are shown. 
        These two optimal paths both contain 2 hops. Comparison between the ApproxNEB energy barriers vs. the relative charge barrier [$\displaystyle \rho_{\rm cyl}(h_{ij}^{\mathbf{K}})/\min\rho_{\rm cyl}(h_{mn}^{\mathbf{K}})$] of all the hops in the five chosen phases of VO(PO$_4$) are shown in (c).
    }
\end{figure}

We examined the total change-density in a radius 1~\AA{} cylinder between sites the end points of a hop; $\rho_{\rm cyl}(h_{ij}^{\mathbf{K}})$.
Since the background charge density can change between different structures, we will only focus on the {\it relative charge barrier}, defined as the ratio between the integrated charge, $\rho_{\rm cyl}(h_{ij}^{\mathbf{K}})$, and its minimum value in that particular structure; $\min(\rho_{\rm cyl}(h_{ij}^{\mathbf{K}}))$.
The relationship between the total charge ratio and the ApproxNEB barrier is shown in Fig.~\ref{fig:aneb} (c), which indicates little correlation between the total charge in the cylinder and the energy barrier. Hence, while promising insertion sites could be identified by low charge-density, it is clear that local atomic relaxations around the working ion during the migration significantly impact the energy barrier such that those effects cannot be ignored. 
However, since the relative charge barrier is an indicator of the amount of negative charge that the migrating ion has to move through we are most interested in migration evens with low relative charge and low ApproxNEB barriers, i.e. the bottom left corner of FIG.~\ref{fig:aneb} (c), for further analysis.

With details of the hops and their connectivity, we can now construct a complete picture of long-range migration in the system. 
In Fig.~\ref{fig:aneb} (a)(b), we show the lowest energy barrier intercalation pathway for two of the structures (mp-25265 and mp-559299) that contain multiple low-barrier hops. In order to reach an accurate description, we performed NEB calculations when evaluating the energy landscape of each hop, the results of which show that both structures contain  an intercalation pathway which has an overall energy barrier of less than 250 eV.

{}

\section{Conclusions}

We demonstrate that the intercalation properties of cations in a solid-state material can be fully captured by a migration graph where the metastable sites represent the nodes and the migration energy barriers are the edge weights.
Using a previously-developed, unbiased cation insertion algorithm, we identify the symmetry-distinct metastable sites in the structure and generate all equivalent sites by repeatedly applying the symmetry operations of the host.
The migration energy is  calculated for the symmetrically-distinct hop between pairs of adjacent metastable sites and the data is replicated on symmetrically equivalent hops to obtain the migration barriers on the entire graph. To identify intercalating pathways, we detect cycles in the periodic graph. Finally, we applied this analysis framework on a diverse set of polymorph structures of VO(PO$_4$) and present several promising structures with low migration barriers.
The framework and code presented here can be used to automatically obtain the migration properties of solid-state materials with essentially no {\it a priori} knowledge. 
Our work opens up opportunities for high throughput studies in the future and can offer a deeper understanding of the migration properties of crystalline solids.

\bibliography{BIBLIO}


\pagebreak
\widetext
\begin{center}
    \textbf{\large Supplemental Materials: Rapid discovery of cathodes, ionic conductors and solid-stateelectrolytes through topological migration analysis}
\end{center}

\setcounter{equation}{0}
\setcounter{figure}{0}
\setcounter{table}{0}
\setcounter{page}{1}
\makeatletter
\renewcommand{\theequation}{S\arabic{equation}}
\renewcommand{\thefigure}{S\arabic{figure}}
\renewcommand{\bibnumfmt}[1]{[S#1]}
\renewcommand{\citenumfont}[1]{S#1}

\section{Migrations graphs for the example materials}

The $\bm{s}_{A/B}$ sites are positions in the host structure that correspond to the Li positions in the inserted structure $S_{A/B}$.
Using the \texttt{SpacegroupAnalyzer} functionality within \texttt{pymatgen} we can apply all of the symmetry transformations of the host structure to the insertions sites.
The position of the inserted site obtained via mapping from the inserted structures, the symmetry transformations of the host as well as resulting transformed position of the inserted site for MnO$_2$ are listed in Table~S\ref{tab:li_positions}.
For CoO$_2$, since all of the allowed symmetry operation of host leaves the inserted site position fixed, they are now listed here.

For MnO$_2$, this results in two symmetry-equivalent copies of $\bm{s}_{A}$ in at $\left(\frac{1}{8},\frac{1}{8},\frac{1}{8}\right)$ and $\left(\frac{7}{8},\frac{7}{8},\frac{7}{8}\right)$ are labeled 0 and 1 respectively.
The four symmetry-equivalent copies of $\bm{s}_{B}$ that form a tetrahedron around $\left(\frac{1}{8},\frac{1}{8},\frac{1}{8}\right)$ are labeled 2 through 5.
The two sites that are symmetrically equivalent to $\bm{s}_{A}$ are labeled 0 and 1, while $\bm{s}_{B}$ sites are labeled 2 through 5.

Using a distance threshold of 3~\AA{}, we find migrations hops between the metastable sites, the hops in the migrations graph for MnO$_2$ are listed in Table~S\ref{tab:mno2_mg} and the hops in the migration graph for CoO$_2$ are listed in Table~S\ref{tab:coo2_mg}.


\newpage

\begin{table*}[h]
    \caption{\label{tab:li_positions}%
        Space group mapping of Li positions $\mathbf{s}_i$ in MnO$_2$ under the space group operations host crystal structure.  The positions of the $s_i$'s and their images are given in fractional coordinates.
    }
    \begin{ruledtabular}
        \begin{tabular}{clccc}
            Index & Label                      & position                                                   & Host transformations\footnote{Determined by the \texttt{SpacegroupAnalyzer}} & Transformed position                                                      \\
            \colrule
            \multirow{4}{*}{0} &\multirow{4}{*}{$\mathbf{s}_A$} & \multirow{4}{*}{$\left(\frac{1}{8},\frac{1}{8},\frac{1}{8}\right)$} & $(x, y, z)$, $(z, -x-y-z+\frac{1}{2}, x)$                                   & \multirow{4}{*}{$\left(\frac{1}{8},\frac{1}{8},\frac{1}{8}\right)$} \\
                                            & &                                                                     & $(-x-y-z+\frac{1}{2}, z, y)$, $(y, x, -x-y-z+\frac{1}{2})$                  &                                                                     \\
                                            & &                                                                     & $(y, z, -x-y-z+\frac{1}{2})$, $(-x-y-z+\frac{1}{2}, x, y)$                  &                                                                     \\
                                            & &                                                                     & $(z, y, x)$, $(x, -x-y-z+\frac{1}{2}, z)$                                   &                                                                     \\
            \colrule
            \multirow{4}{*}{1} &\multirow{4}{*}{$\mathbf{s}_A$} & \multirow{4}{*}{$\left(\frac{1}{8},\frac{1}{8},\frac{1}{8}\right)$} & $(-y, -z, x+y+z+\frac{1}{2})$, $(x+y+z+\frac{1}{2}, -x, -y)$                & \multirow{4}{*}{$\left(\frac{7}{8},\frac{7}{8},\frac{7}{8}\right)$} \\
                                            & &                                                                     & $(-z, -y, -x)$, $(-x, x+y+z+\frac{1}{2}, -z)$                               &                                                                     \\
                                            & &                                                                     & $(-x, -y, -z)$, $(-z, x+y+z+\frac{1}{2}, -x)$                               &                                                                     \\
                                            & &                                                                     & $(x+y+z+\frac{1}{2}, -z, -y)$, $(-y, -x, x+y+z+\frac{1}{2})$                &                                                                     \\
            \colrule
            \multirow{2}{*}{4} &\multirow{2}{*}{$\mathbf{s}_B$} & \multirow{2}{*}{$\left(0,0,0\right)$}                               & $(x, y, z)$, $(-z, -y, -x)$,                                                & \multirow{2}{*}{$\left(0,0,0\right)$}                               \\
                                            & &                                                                     & $(-x, -y, -z)$, $(z, y, x)$                                                 &                                                                     \\
            \colrule
            \multirow{2}{*}{3} &\multirow{2}{*}{$\mathbf{s}_B$} & \multirow{2}{*}{$\left(0,0,0\right)$}                               & $(x+y+z+\frac{1}{2}, -x, -y)$, $(-x-y-z+\frac{1}{2}, z, y)$                 & \multirow{2}{*}{$\left(\frac{1}{2},0,0\right)$}                     \\
                                            & &                                                                     & $(-x-y-z+\frac{1}{2}, x, y)$, $(x+y+z+\frac{1}{2}, -z, -y)$                 &                                                                     \\
            \colrule
            \multirow{2}{*}{2} &\multirow{2}{*}{$\mathbf{s}_B$} & \multirow{2}{*}{$\left(0,0,0\right)$}                               & $(z, -x-y-z+\frac{1}{2}, x)$, $(-x, x+y+z+\frac{1}{2}, -z)$                 & \multirow{2}{*}{$\left(0,\frac{1}{2},0\right)$}                     \\
                                            & &                                                                     & $(-z, x+y+z+\frac{1}{2}, -x)$, $(x, -x-y-z+\frac{1}{2}, z)$                 &                                                                     \\
            \colrule
            \multirow{2}{*}{5} & \multirow{2}{*}{$\mathbf{s}_B$} & \multirow{2}{*}{$\left(0,0,0\right)$}                               & $(-y, -z, x+y+z+\frac{1}{2})$, $(y, x, -x-y-z+\frac{1}{2})$,                & \multirow{2}{*}{$\left(0,0,\frac{1}{2}\right)$}                     \\
                                            & &                                                                     & $(y, z, -x-y-z+\frac{1}{2})$, $(-y, -x, x+y+z+\frac{1}{2})$                 &                                                                     \\
        \end{tabular}
    \end{ruledtabular}
\end{table*}

\begin{table}
    \caption{\label{tab:mno2_mg}%
        Full list of the migration events ($h_{ij}^{\mathbf{K}}$) in MnO$_2$ that are less than 3~\AA{} and are distinct under lattice-vector translations.
        The hops with the same label are equivalent under some space group operation.
    }
    \begin{ruledtabular}
        \begin{tabular}{llrrr}
            $i$-index & $j$-index & $j$-image vector ($\mathbf{K}$) & distance & label \\
            \colrule
            0 & 2 & (0, 0, 0) & 1.784 & 0 \\
            0 & 3 & (0, 0, 0) & 1.784 & 0 \\
            0 & 4 & (0, 0, 0) & 1.784 & 0 \\
            0 & 5 & (0, 0, 0) & 1.784 & 0 \\
            1 & 2 & (1, 1, 1) & 1.784 & 0 \\
            1 & 3 & (0, 1, 1) & 1.784 & 0 \\
            1 & 4 & (1, 0, 1) & 1.784 & 0 \\
            1 & 5 & (1, 1, 0) & 1.784 & 0 \\
            2 & 3 & (0, 0, 0) & 2.913 & 1 \\
            2 & 3 & (-1, 0, 0) & 2.913 & 1 \\
            2 & 4 & (0, 0, 0) & 2.913 & 1 \\
            2 & 4 & (0, -1, 0) & 2.913 & 1 \\
            2 & 5 & (0, 0, 0) & 2.913 & 1 \\
            2 & 5 & (0, 0, -1) & 2.913 & 1 \\
            3 & 4 & (0, 0, 0) & 2.913 & 1 \\
            3 & 4 & (1, -1, 0) & 2.913 & 1 \\
            3 & 5 & (0, 0, 0) & 2.913 & 1 \\
            3 & 5 & (1, 0, -1) & 2.913 & 1 \\
            4 & 5 & (0, 0, 0) & 2.913 & 1 \\
            4 & 5 & (0, 1, -1) & 2.913 & 1 \\
        \end{tabular}
    \end{ruledtabular}
\end{table}

\begin{table}
    \caption{\label{tab:coo2_mg}%
        Full list of the migration events ($h_{ij}^{\mathbf{K}}$) in CoO$_2$ that are less than 3~\AA{} and are distinct under lattice-vector translations.
        The hops with the same label are equivalent under some space group operation.
    }
    \begin{ruledtabular}
        \begin{tabular}{llrrr}
            $i$-index & $j$-index & $j$-image vector ($\mathbf{K}$) & distance & label \\
            \colrule
            0         & 1        & (-1, 0, 0)                      & 2.846    & 0     \\
            0         & 1        & (-1, 1, 0)                      & 2.846    & 0     \\
            0         & 1        & (0, 0, 0)                       & 2.846    & 0     \\
            0         & 1        & (0, 1, 0)                  & 2.846    & 0     \\
            0         & 0        & (-1, 0, 0)                      & 2.820    & 1     \\
            0         & 0         & (1, 0, 0)                       & 2.820    & 1     \\
            1         & 1         & (-1, 0, 0)                      & 2.820    & 2     \\
            1         & 1         & (1, 0, 0)                       & 2.820    & 2     \\
        \end{tabular}
    \end{ruledtabular}
\end{table}

\clearpage

\section{Intercalating pathway finding}

The charge density analysis developed for the approximate NEB workflow~\cite{Rong2016Aug} calculates an optimal pathway between two point using the electronic charge density as a ``virtual'' potential.
We assign a charge barrier $\rho_{b}$ to a given unique hop as the peak averages charge density (in a sphere of 0.4~\AA) along the path.
Using a simple cost function of the charge barrier times the distance of the hop, we can assign a cost to each hop or edge of our migration graph.

\begin{table}[h]
    \centering
    \caption{ApproxNEB energy barrier and charge barrier of mp-1104567}
    \begin{ruledtabular}
        \begin{tabular}{crrr}
            $\bm{r}_i \rightarrow \bm{r}_j$                  & Barrier (eV) & $\rho^{\rm cyl}_{i,j}$ (e$^{-}$) & $\rho^{max}_{i,j}$ (e$^{-}$/\AA{}$^3$)\\
            \colrule
            $(0.27, 0.73, 0.0)\rightarrow(0.5, 0.5, 0.5)$    & 0.370        & 0.050                            & 2.992          \\
            $(0.27, 0.73, 0.0)\rightarrow(0.73, 0.27, 0.0)$  & 0.434        & 0.012                            & 0.576          \\
            $(0.27, 0.73, 0.0)\rightarrow(-0.27, 1.27, 0.0)$ & 0.769        & 0.019                            & 0.541          \\
        \end{tabular}
    \end{ruledtabular}
\end{table}

\begin{table}[h]
    \centering
    \caption{ApproxNEB energy barrier and charge barrier of mp-556459}
    \begin{ruledtabular}
        \begin{tabular}{crrr}
            $\bm{r}_i \rightarrow \bm{r}_j$                    & Barrier (eV) & $\rho^{\rm cyl}_{i,j}$ (e$^{-}$) & $\rho^{max}_{i,j}$ (e$^{-}$/\AA{}$^3$)\\
            \colrule
            $(0.5, 0.5, 0.5)\rightarrow(0.5, 1.0, 0.0)$        & 2.340        & 0.025                            & 1.617          \\
            $(0.0, 0.5, 0.5)\rightarrow(-0.32, 0.32, 0.75)$    & 2.437        & 0.013                            & 2.813          \\
            $(0.0, 0.5, 0.5)\rightarrow(-0.5, 0.0, 1.0)$       & 2.445        & 0.073                            & 4.324          \\
            $(0.32, 0.68, 0.25)\rightarrow(-0.32, 0.32, 0.75)$ & 2.473        & 0.026                            & 2.813          \\
            $(0.5, 0.5, 0.5)\rightarrow(0.32, 0.68, 0.25)$     & 2.514        & 0.006                            & 2.586          \\
            $(0.5, 0.5, 0.5)\rightarrow(0.0, 0.5, 0.5)$        & 2.602        & 0.027                            & 1.955          \\
            $(0.5, 0.5, 0.5)\rightarrow(0.5, 0.5, 0.0)$        & 3.033        & 0.004                            & 0.162          \\
            $(0.5, 0.5, 0.5)\rightarrow(1.32, 0.68, 0.25)$     & 3.071        & 0.046                            & 1.874          \\
        \end{tabular}
    \end{ruledtabular}
\end{table}

\begin{table}[h]
    \centering
    \caption{ApproxNEB energy barrier and charge barrier of mp-25265}
    \begin{ruledtabular}
        \begin{tabular}{crrr}
            $\bm{r}_i \rightarrow \bm{r}_j$                     & Barrier (eV) & $\rho^{\rm cyl}_{i,j}$ (e$^{-}$) & $\rho^{max}_{i,j}$ (e$^{-}$/\AA{}$^3$)\\
            \colrule
            $(0.5, 0.5, 0.0)\rightarrow(0.0, 0.5, 0.0) $        & 0.211        & 0.002                            & 0.106          \\
            $(0.5, 0.5, 0.0)\rightarrow(0.25, 0.43, -0.03) $    & 0.285        & 0.001                            & 0.048          \\
            $(0.25, 0.43, 0.97)\rightarrow(-0.25, 0.57, 1.03) $ & 0.336        & 0.002                            & 0.053          \\
        \end{tabular}
    \end{ruledtabular}
\end{table}

\begin{table}[h]
    \centering
    \caption{ApproxNEB energy barrier and charge barrier of mp-559299}
    \begin{ruledtabular}
        \begin{tabular}{crrr}
            $\bm{r}_i \rightarrow \bm{r}_j$                   & Barrier (eV) & $\rho^{\rm cyl}_{i,j}$ (e$^{-}$) & $\rho^{max}_{i,j}$ (e$^{-}$/\AA{}$^3$)\\
            \colrule
            $(0.25, 0.75, 0.0)\rightarrow(0.0, 1.0, 0.0) $    & 0.200        & 0.008                            & 0.494          \\
            $(0.25, 0.75, 0.0)\rightarrow(-0.25, 0.75, 0.0) $ & 0.326        & 0.015                            & 0.570          \\
        \end{tabular}
    \end{ruledtabular}
\end{table}

\begin{table}[h]
    \centering
    \caption{ApproxNEB energy barrier and charge barrier of mp-763482}
    \begin{ruledtabular}
        \begin{tabular}{crrr}
            $\bm{r}_i \rightarrow \bm{r}_j$                     & Barrier (eV) & $\rho_{\rm tot}$ (e$^{-}$) & $\rho^{max}_{i,j}$ (e$^{-}$/\AA{}$^3$)\\
            \colrule
            $(0.24, 0.82, 0.07)\rightarrow(0.26, 0.68, -0.07) $ & 0.005        & 0.005                      & 2.985          \\
            $(0.24, 0.82, 0.07)\rightarrow(0.5, 0.5, 0.0) $     & 0.138        & 0.041                      & 3.136          \\
            $(0.24, 0.82, 0.07)\rightarrow(0.18, 1.24, -0.07) $ & 0.184        & 0.057                      & 2.660          \\
            $(0.24, 0.82, 0.07)\rightarrow(0.0, 1.0, 0.0) $     & 0.193        & 0.010                      & 3.135          \\
            $(0.24, 0.82, 0.07)\rightarrow(-0.32, 0.74, 0.07) $ & 0.196        & 0.042                      & 2.209          \\
            $(0.24, 0.82, 0.07)\rightarrow(0.18, 0.24, -0.07) $ & 0.203        & 0.042                      & 3.134          \\
            $(0.24, 0.82, 0.07)\rightarrow(-0.24, 1.18, 0.07) $ & 0.203        & 0.016                      & 3.134          \\
            $(0.24, 0.82, 0.07)\rightarrow(0.68, 0.74, 0.07) $  & 0.227        & 0.063                      & 2.257          \\
            $(0.24, 0.82, 0.07)\rightarrow(0.76, 1.18, 0.07) $  & 0.231        & 0.027                      & 3.134          \\
            $(0.24, 0.82, 0.07)\rightarrow(0.26, 0.68, 0.93) $  & 0.930        & 0.157                      & 4.883          \\
        \end{tabular}
    \end{ruledtabular}
\end{table}

\newpage

\begin{algorithm}[H]
    \caption{Dijkstra's algorithm for a periodic graph}
    \label{algo:periodic}
     \textbf{Inputs}: \\
     \hspace*{\algorithmicindent} $cost[i,j,K]$ --- The cost data associated with hop $h_{i,j}^{\bm{K}}$ \\
     \hspace*{\algorithmicindent} $u$ ---  Starting node index \\  
    \textbf{Outputs}: \\ 
    \hspace*{\algorithmicindent} The minimum cost of from node $u$ to itself with finite cumulitive displacement.
    \begin{algorithmic}[1]
        \State $minCost[v, D]$ $\leftarrow$ Mapping with default value INFINITY
        \State $prev[v, D]$ $\leftarrow$ Mapping with default value NULL
        \State $minCost[u, D=(0,0,0)]$ $\leftarrow$ Set to 0
        \State $Q$ $\leftarrow$ add $(v, (0,0,0))$ to a queue
        \While{$Q$ is not empty}
        \State $u, D$ $\leftarrow$ pop vertex in $Q$ with lowest cost
        \For{each ($v$, $K$) neighbor of $u$}

        \If{$u < v$ \textbf{or} ($u==v$ \textbf{and} first non-zero index of $K$ is positive)}
        \State $D^\prime \leftarrow D + K$
        \Else{}
        \State $D^\prime \leftarrow D - K$
        \EndIf

        \State $newCost$ = $minCost[u,D]$ + $cost[u,v,K]$
        
        \If{$D^\prime$ within some user-defined limit \textbf{and} $newCost < minCost[v, D^\prime]$}
            \State $minCost[v, D^\prime] = newCost$ 
            \State $prev[v, D^\prime] = (u, D)$ 
        \EndIf
        \EndFor
        \EndWhile
        \State \Return $cost[u, D]$ where $D$ not (0,0,0)

    \end{algorithmic}
\end{algorithm}


\end{document}